\documentclass[onecolumn,aps,prd,preprintnumbers,superscriptaddress,nofootinbib,amsmath,amssymb,floats,floatfix,showkeys,notitlepage,showpacs]{revtex4-2}

\usepackage{orcidlink}
\usepackage{comment}
\usepackage{lipsum}
\usepackage{graphicx}
\usepackage{subfigure}
\usepackage{palatino}
\usepackage{sans}
\usepackage{hyperref}
\hypersetup{colorlinks=true,linkcolor=blue,urlcolor=blue,citecolor=blue}
\usepackage[toc,page]{appendix}
\usepackage[normalem]{ulem}
\usepackage{adjustbox}
\usepackage{latexsym}
\usepackage{amsmath}
\usepackage{amssymb}
\usepackage{amsfonts}
\usepackage{dcolumn}
\usepackage{bm}
\usepackage{tikz}
\usepackage{bigints}
\usepackage{array,tabularx,multirow,booktabs}
\usepackage[tracking=true]{microtype}
\SetTracking{}{500}
\SetTracking{encoding={*}, shape=sc}{40}
\UseRawInputEncoding 
\allowdisplaybreaks

\begin{document} \sloppy
\title{Traces of quantum gravitational correction at third-order curvature through the black hole shadow and particle deflection at the weak field limit}

\author{Gaetano Lambiase \orcidlink{0000-0001-7574-2330}}
\email{lambiase@sa.infn.it}
\affiliation{Dipartimento di Fisica ``E.R Caianiello'', Università degli Studi di Salerno, Via Giovanni Paolo II, 132 - 84084 Fisciano (SA), Italy.}
\affiliation{Istituto Nazionale di Fisica Nucleare - Gruppo Collegato di Salerno - Sezione di Napoli, Via Giovanni Paolo II, 132 - 84084 Fisciano (SA), Italy.}

\author{Reggie C. Pantig
\orcidlink{0000-0002-3101-8591}
}
\email{rcpantig@mapua.edu.ph}
\affiliation{Physics Department, Map\'ua University, 658 Muralla St., Intramuros, Manila 1002, Philippines}

\author{Ali \"Ovg\"un
\orcidlink{0000-0002-9889-342X}
}
\email{ali.ovgun@emu.edu.tr}
\affiliation{Physics Department, Eastern Mediterranean
University, 99628, Famagusta, North Cyprus
via Mersin 10, Turkiye.}

\begin{abstract}
This study investigates the impact of the quantum-gravity correction at the third-order curvature ($c_6$) on the black hole's shadow and deflection angle on the weak field regime, both involving finite distances of observers. While the calculation of the photonsphere and shadow radius $R_{\rm sh}$ can easily be achieved by the standard Lagrangian for photons, the deflection angle $\alpha$ employs the finite-distance version of the Gauss-Bonnet theorem (GBT). We find that the photonsphere reduces to the classical expression $r_{\rm ph} = 3M$ for both the Planck mass and the theoretical mass limit for BH, thus concealing the information about the applicability of the metric on the quantum and astrophysical grounds. Our calculation of the shadow, however, revealed that $c_6$ is strictly negative and constrains the applicability of the metric to quantum black holes. For instance, the bounds for the mass is $M/l_{\rm Pl} \in [0.192, 4.315]$. We also derived the analytic formula for the observer-dependent shadow, which confirms $c_6$'s influence on quantum black holes even for observers in the asymptotic regions. The influence of such a parameter also strengthens near the quantum black hole. Our analytic calculation of $\alpha$ is shown to be independent of $c_6$ if the finite distance $u \rightarrow 0$, and $c_6$ is not coupled to any time-like geodesic. Finally, the effect of $c_6$ manifests in two ways: if $M^2$ is large enough to offset the small value of $l_{\rm Pl}$ (which is beyond the theoretical mass limit), or if $b$ is comparable to $l_{\rm Pl}$ for a quantum black hole.
\end{abstract}


\maketitle


\section{Introduction} \label{intr}
Black holes hold profound scientific interest due to their multifaceted nature. These objects have captivated astronomers for decades, not only due to their extreme properties but also because they provide unique laboratories for testing the limits of our current understanding of physics, particularly in general relativity and quantum mechanics. However, these two realms proved difficult to reconcile as they fundamentally contradict each other near black hole singularities, where gravity becomes infinitely strong. Quantum corrections attempt to bridge this gap by incorporating quantum effects into the description of gravity, potentially resolving this paradox. We lack a complete theory of quantum gravity, the unification of general relativity and quantum mechanics. Studying quantum corrections in black holes provides a unique testing ground for various proposed quantum gravity theories, helping us distinguish between them and potentially leading to a breakthrough in our understanding of gravity at its most fundamental level \cite{Calmet:2017qqa,Calmet:2018elv,Calmet:2021lny,Calmet:2023gbw,Kiefer:2020cbu,Binetti:2022xdi,Mertens:2022irh,DelPiano:2023fiw,Devi:2021ctm}.

One of the goals of this theoretical work is to study the quantum-corrected Schwarzschild black hole by examining the imprints or traces of the quantum gravity corrections to the geometry of the shadow and uncover what restrictions it can bring. This is important since the black hole shadow offers a direct, visual probe of its geometry, whether at the quantum or astrophysical level. By measuring its size, shape, and distortions, we can constrain the black hole's mass, spin, and even the presence of surrounding material like accretion disks. Comparing the observed shadow with predictions from General Relativity (GR) reveals any deviations, potentially hinting at alternative theories of gravity or quantum corrections to its predictions \cite{Saadati:2023jym, Jiang:2023img,Lambiase:2023hng,Atamurotov:2022iwj,Yang:2022btw, Tang:2022uwi,Lobos:2022jsz,Xu:2021xgw,Devi:2021ctm,Zakharov:2021gbg,Zakharov:2014lqa,Vagnozzi:2022moj,Vagnozzi:2019apd,Allahyari:2019jqz,Khodadi:2020jij,Meng:2023htc,Li:2023djs,Kuang:2022ojj,Abdujabbarov:2016hnw,Atamurotov:2013sca,Atamurotov:2015xfa,Islam:2020xmy,Kumar:2020owy,Afrin:2021imp,Tsukamoto:2014tja,Tsukamoto:2017fxq,Ovgun:2023wmc}. The optical manifestation of black holes has been under investigation since the pioneering work of Cunningham and Bardeen \cite{Cunningham} and Bardeen \cite{1974IAUS...64..132B}, who initially explored the dynamics of a star in orbit around a black hole, alongside other relevant scenarios. Their early investigations delineated the essential configuration of a slender disk featuring a centrally located area subjected to gravitational lensing effects. Luminet (1979) \cite{Luminet:1979nyg} subsequently presented the initial computer-generated representation of a black hole encompassed by a luminous accretion disk, albeit in a hand-drawn format. The term "shadow" was independently introduced shortly afterward by Falcke et al. \cite{Falcke:1999pj} and de Vries \cite{deVries:1999tiy}. Subsequent to its inception, the Event Horizon Telescope (EHT) has adopted this nomenclature to elucidate the observable phenomenon in black hole imaging. Building on their groundbreaking 2017 image of M87*, the supermassive black hole at the heart of galaxy Messier 87, the Event Horizon Telescope (EHT) Collaboration captured new images in 2018 \cite{EventHorizonTelescope:2019dse,EventHorizonTelescope:2019ths, EventHorizonTelescope:2022xqj}. These latest observations confirm the familiar ring of emission around a central dark zone, consistent with the "shadow" cast by the black hole as predicted by general relativity. Further studies aimed at Sagittarius A*, the black hole at the center of our own Milky Way, employed extensive simulations and revealed its appearance aligns with theoretical models of a Kerr black hole \cite{EventHorizonTelescope:2022wkp,EventHorizonTelescope:2022wok}.These achievements open exciting avenues for exploring black hole astrophysics and rigorously testing the fundamental principles of general relativity through direct observation. Hence, we cannot underestimate theoretical efforts since more sophisticated technologies and methods might reveal the secrets of a black hole.

Another aim of the paper is to additionally probe the effects of the quantum gravity correction on the deflection angle of null and timelike particles in the weak field limit. Quantum corrections, arising from theories like quantum gravity, may introduce subtle deviations in the gravitational lensing signatures in the weak field regime Thus, by analyzing these observational phenomena, we can potentially uncover evidence for quantum gravitational effects and refine our understanding of the fundamental nature of black holes and spacetime at the quantum level. Gravitational lensing (GL) manifests as the deflection of light from distant sources by intervening massive objects like galaxies or black holes. This phenomenon exhibits distinct weak and strong field regimes. Weak GL involves slight light path bending, leading to distorted images of distant objects, and has sparked significant research in astrophysics and cosmology. The seminal 1919 Eddington expedition confirmed Einstein's theory of relativity through GL observations. Determining object distances is crucial for understanding their properties; however, Virbhadra \cite{Virbhadra:2022ybp} demonstrated that solely analyzing relativistic images, without mass and distance information, can accurately bound the compactness of massive dark objects. Furthermore, Virbhadra identified a distortion parameter that renders the signed sum of all singular GL images to vanish (tested with Schwarzschild lenses in both weak and strong fields, \cite{Virbhadra:2022iiy}). Several methods have been developed over the years to calculate the deflection angle caused by black holes and other compact objects in asymptotically/non-asymptotically flat spacetimes \cite{Virbhadra:1999nm,Virbhadra:2002ju,Virbhadra:2008ws,Nascimento:2020ime,Furtado:2020puz,kuang2022strong}. Gibbons and Werner (2008) offered a distinct approach to calculate the deflection angle in weak fields using the Gauss-Bonnet theorem (GBT) on asymptotically flat spacetimes' optical geometries \cite{Gibbons:2008rj}. This requires integrating the GBT across an infinite domain bounded by the light ray. Werner (2013) extended the GBT deflection angle technique to stationary spacetimes within the Finsler-Randers type optical geometry on Nazim's osculating Riemannian manifolds \cite{Werner_2012}. Subsequently, Ishihara et al. (2015, 2016) devised a method to apply the GBT differently to finite distances, departing from the traditional approach with asymptotic receiver and source \cite{Ishihara:2016vdc,Ishihara:2016sfv}. This methodology was further extended by Ono et al. (2016) to axially symmetric spacetimes \cite{Ono:2017pie} and applied to various non-asymptotic spacetimes, including those with dark matter contributions \cite{Pantig:2022toh,Pantig:2022whj,Ovgun:2023ego}. Finally, Li et al. (2018, 2019) explored the finite distance method using massive particles and the Jacobi-Maupertuis Randers-Finsler metric within the GBT framework \cite{Li:2020dln,Li:2020wvn}.

Here is the program of the paper: Sect. \ref{sec2} emphasizes the brief introduction of the Schwarzschild metric with a quantum gravitation correction at the third-order curvature. We study the photon-sphere in Sect. \ref{sec3} and attempt to derive an analytical formula for the observer-dependent shadow radius in Sect. \ref{sec4}. Then, we calculate the weak deflection angle in Sect. \ref{sec5}. Finally, we form conclusive remarks in Sect. \ref{conc}. We use the metric signature $(-,+,+,+)$ and the natural units $G = c = \bar{h} = 1$ throughout the analysis in the paper.

\section{Brief Review of Schwarzschild black hole with quantum-correction at third order curvature} \label{sec2}
Calmet and Kuipers recently employed effective field theory to compute quantum gravitational entropy corrections for black holes, utilizing the Wald entropy formula \cite{Wald:1993nt}. Their work unveiled intriguing connections between quantum entropy corrections, Euler characteristic, and quantum metric corrections for Schwarzschild black holes. They proposed a systematic approach to any perturbation order or black hole metric. The Wald entropy formula, as defined by \cite{Wald:1993nt}, is expressed as:
\begin{eqnarray}
S_{Wald} =  -2\pi \int d\Sigma \  \epsilon_{\mu\nu} \epsilon_{\rho\sigma}
	 \frac{\partial \mathcal{L}}{\partial R_{\mu\nu\rho\sigma}}\Big|_{r=r_H} ,
\end{eqnarray}
In this context, $d\Sigma=r^2 \sin \theta d\theta d\phi$, $\mathcal{L}$ denotes the Lagrangian of the model, $R^{\mu\nu\rho\sigma}$ stands for the Riemann tensor, and $r_H$ represents the horizon radius. Additionally, it holds that  $\epsilon_{\mu\nu}\epsilon^{\mu\nu}=-2$, $\epsilon_{\mu\nu}=-\epsilon_{\nu\mu}$. 

When employing the effective action in quantum gravity \cite{Donoghue:1994dn}, at the second order in curvature, one finds:
\begin{align}\label{EFTaction}
S_{\text{EFT}} = \int \sqrt{|g|}d^4x  \left( \frac{R}{16\pi} + c_1(\mu) R^2 + c_2(\mu) R_{\mu\nu} R^{\mu\nu} + c_3(\mu) R_{\mu\nu\rho\sigma} R^{\mu\nu\rho\sigma} + \mathcal{L}_m \right) \ .
\end{align}
The non-local part of the action is given by
\begin{align}\label{nonlocalaction}
	\Gamma_{\text{NL}}^{\scriptstyle{(2)}}  = - \int  \sqrt{|g|}d^4x \left[ \alpha R \ln\left(\frac{\Box}{\mu^2}\right)R + \beta R_{\mu\nu} \ln\left(\frac{\Box}{\mu^2}\right) R^{\mu\nu} + \gamma R_{\mu\nu\alpha\beta} \ln\left(\frac{\Box}{\mu^2}\right)R^{\mu\nu\alpha\beta} \right],
\end{align}
where $\Box := g^{\mu\nu} \nabla_\mu \nabla_\nu$. To show that there are no corrections up to second order in curvature to the Schwarzschild metric is straightforward if one uses the non-local Gauss-Bonnet identity, as shown in \cite{Calmet:2017qqa,Calmet:2018elv}
\begin{eqnarray}
\int   \sqrt{|g|} d^4x R_{\mu\nu\alpha\beta}\left (c_3(\mu)-\gamma  \ln\left(\frac{\Box}{\mu^2}\right) \right) R^{\mu\nu\alpha\beta} &=&
+4 \int  \sqrt{|g|} d^4xR_{\mu\nu}\left (c_3(\mu)-\gamma  \ln\left(\frac{\Box}{\mu^2}\right)\right)R^{\mu\nu}  \nonumber \\ &&
-\int   \sqrt{|g|} d^4xR\left (c_3(\mu)-\gamma  \ln\left(\frac{\Box}{\mu^2}\right)\right)R \nonumber \\ && + {\cal O}(R^3)+ {\rm  boundary \ terms.}
\end{eqnarray}
This identity can be established with reference to works such as \cite{Knorr:2019atm,Donoghue:2015nba}
\begin{eqnarray}
\log \frac{\Box}{\mu^2}=\int_0^\infty ds \frac{e^{-s}-e^{-s\frac{\Box}{\mu^2}} }{s}
\end{eqnarray}
and  \cite{Barvinsky:1990up}
\begin{eqnarray}
\Box R^{\alpha\beta\mu\nu}&=&\nabla^\mu\nabla^\alpha R^{\nu\beta} -\nabla^\nu\nabla^\alpha R^{\mu\beta}-\nabla^\mu\nabla^\beta R^{\nu\alpha}+
+\nabla^\nu\nabla^\beta R^{\mu\alpha}\nonumber \\ && -4  R^{\alpha \ [\mu}_{\ \sigma \ \lambda} R^{\beta\sigma\nu]\lambda}
+ 2 R^{ [ \mu}_{\ \lambda} R^{\alpha\beta\lambda\nu] }- R^{\alpha\beta}_{\ \ \sigma \lambda}R^{\mu\nu\sigma\lambda}.
 \end{eqnarray}
 This result follows from the Bianchi identity. One obtains\cite{Knorr:2019atm,Deser:1986xr,Asorey:1996hz,Teixeira:2020kew}
\begin{eqnarray}
R_{\alpha\beta\mu\nu}\Box R^{\alpha\beta\mu\nu}= 4 R_{\alpha\beta\mu\nu}\nabla^\alpha \nabla^\mu R^{\beta\nu}+ {\cal O}(R^3).
 \end{eqnarray}
 
Generalizing this result to higher powers of the Laplacian is straightforward. We derive the non-local Gauss-Bonnet identity by inserting this relation into the Lagrangian and employing partial integrations along with the contracted Bianchi identity. At the second order in curvature, it is found that the Riemann tensor can be eliminated from the dynamical part of the action. Consequently, there are no corrections to the field equations at this order for vacuum solutions of general relativity, as highlighted in \cite{Calmet:2018elv}.
\begin{eqnarray} \label{2ndentropy}
S_{Wald}^{(2)}  &=&\frac{A}{4} +  64 \pi^2 c_3(\mu) +64 \pi^2 \gamma \left( \log \left (4 M^2 \mu^2\right) -2 +2 \gamma_E \right)
\end{eqnarray}
Here, $A=16 \pi M^2$ represents the area of the black hole. A similar result was attained using the Euclidean path integral formulation. It's important to note that the entropy remains renormalization group invariant and finite. Given the absence of metric corrections, the temperature remains unchanged, and the classical relation $TdS=dM$ undergoes a quantum correction. Specifically, we find $T dS= (1+\gamma 16 \pi /( M^2))dM$.

The first law of thermodynamics is then expressed as:
\begin{eqnarray}
TdS-PdV=\left(1+\gamma \frac{16 \pi}{ M^2}\right) dM=  dM+\gamma \frac{16 \pi}{ M^2} dM.
\end{eqnarray}
Here, $P$ denotes the pressure of the black hole. Its volume is defined as $V=\frac{4}{3} \pi r_H^3$, where $r_H= 2  M$ represents the horizon radius. One can then establish the following identification: $TdS=dM$ and $\gamma 16 \pi/( M^2) dM= - P dV$  with $dV=32 \pi  M^2 dM$. One obtains
\begin{eqnarray} \label{pressure}
P=-\gamma \frac{1}{2 M^4}.
\end{eqnarray}
The pressure $P$ can be negative if $\gamma$ is positive for spin-0, spin-1/2, and spin-2 fields or positive if $\gamma$ is negative for spin-1 fields. This is confirmed by finding:$\gamma_0=2/(11520 \pi^2)$ \cite{Deser:1974cz},  $\gamma_{1/2}=7/(11520 \pi^2)$ \cite{Deser:1974cz}, $\gamma_{1}=-26/(11520 \pi^2)$ \cite{Deser:1974cz} and $\gamma_{2}=424/(11520 \pi^2)$ \cite{Barvinsky:1984jd}.  Dolan (2011) explored the idea of black holes having pressure within gravitational models incorporating a cosmological constant. Quantum gravity reveals pressure for Schwarzschild black holes, a departure from prior works that overlooked quantum metric corrections \cite{Fursaev:1994te,El-Menoufi:2015cqw,El-Menoufi:2017kew}. Since there are no dynamical metric corrections at this curvature order, interpreting the entropy correction as a pressure term is necessary.

At the third order in curvature, additional operators need to be incorporated into the effective action
\begin{eqnarray}
{\cal L}^{(3)}=c_6  R^{\mu\nu}_{\ \ \alpha\sigma} R^{\alpha\sigma}_{\ \ \delta\gamma} R^{\delta\gamma}_{\ \ \mu\nu}  \ ,
\end{eqnarray}
As highlighted by Goroff and Sagnotti \cite{Goroff:1985th}, in vacuum, there exists only one invariant comprising solely Riemann tensors, given by: $R_{\alpha\beta\gamma\delta} R^{\alpha \ \gamma}_{\ \epsilon \ \zeta} R^{\beta\epsilon\delta\zeta}$ in terms of $R^{\mu\nu}_{\ \ \alpha\sigma} R^{\alpha\sigma}_{\ \ \delta\gamma} R^{\delta\gamma}_{\ \ \mu\nu}$ Additionally, terms involving the Ricci scalar or Ricci tensors vanish in vacuum. There exists a corresponding non-local operator $R^{\mu\nu}_{\ \ \alpha\sigma} \log{\Box} R^{\alpha\sigma}_{\ \ \delta\gamma}R^{\delta\gamma}_{\ \ \mu\nu}$. Although the Wilson coefficient is known in a specific gauge \cite{Goroff:1985th}, it remains unknown for the unique effective action. Therefore, we will neglect this term.

In \cite{Calmet:2017qqa,Calmet:2018elv,Calmet:2021lny,Calmet:2023gbw}, it has been demonstrated that the leading-order quantum gravitational correction to the Schwarzschild metric emerges at the third order in curvature. This correction is generated by a local operator $c_6  R^{\mu\nu}_{\ \ \alpha \sigma} R^{\alpha\sigma}_{\ \ \delta \gamma} R^{\delta \gamma}_{\ \ \mu\nu}$. Since this operator is local, its Wilson coefficient cannot be computed from the first principles within this approach. Instead, it must be determined from experimental observations. The leading-order quantum gravitational correction to the Schwarzschild metric is \cite{Calmet:2021lny}
\begin{align}
    ds^2&=-A(r)dt^2+B(r)dr^2+r^2d\Omega^2, \nonumber \\
    A(r)&=1-\frac{2 G M}{r}+640\pi c_6 \frac{ G^5 M^3}{r^7} \nonumber \\
    B(r)&=\left[1-\frac{2 G M}{r}+128\pi c_6 \frac{ G^4 M^2}{r^6}\left(27-\frac{ 49G M}{r}\right) \right]^{-1},
\end{align}
where $d\Omega^2=d\theta^2+r^2\sin^2\theta$  denotes the line element of the unit two-spheres. As a final note on these metric functions, we must ensure they are dimensionless if we are working with black holes for the sake of brevity. For instance, recalling Planck units, $G = M_{\rm Pl}^{-2}$, where $M_{\rm Pl}$ is the Planck mass equal to $2.176434\times 10^{-8}$ kg. However, this makes the metric function $A(r)$ have a dimension of $([m][kg])^{-1}$ on its second term. If we ought to geometrize both the black hole mass and the Planck mass, the above metric functions will transform to
\begin{align} \label{eq:gqs}
    A(r)&=1-\frac{2 M}{r}+640\pi c_6 \frac{ l_{\rm Pl}^4 M^3}{r^7}, \nonumber \\
    B(r)&=\left[1-\frac{2 M}{r}+128\pi c_6 \frac{ l_{\rm Pl}^4 M^2}{r^6}\left(27-\frac{ 49 M}{r}\right) \right]^{-1},
\end{align}
where $l_{\rm Pl} = 1.616255 \times 10^{-35}$ m is the Planck length. Inspecting these expressions, it can be argued that if $M$ represents quantum or astrophysical black hole masses (constrained by both observation and theoretical limits), we see that Eq. \eqref{eq:gqs} reduces immediately to the Schwarzschild case. An exception occurs if $M$ is so large that it offsets the small value of $l_{\rm Pl}$. Thus, this work aims to explore the black hole phenomenon, which may confirm the applicability of this metric at the quantum level.



\section{Calculations of photon-sphere radius} \label{sec3}

This section explores the photon-sphere due to the quantum correction at the third-order curvature. Considering the metric coefficients in Eq. \eqref{eq:gqs}, we examine the Lagrangian $\mathcal{L} (x, \dot{x} ) = (1/2) g_{\mu \nu} \dot{x}^{\mu} \dot{x}^{\nu}$ for null geodesics
\begin{equation}
    \mathcal{L} (x , \dot{x} ) \, = \, 
\dfrac{1}{2} \, \left(- \, A(r) \,   \, \dot{t}{}^2
+ B(r) \, \dot{r}^2 + r^2 ( \dot{\theta}^2 + \sin^2 \theta \, \dot{\phi}^2 )  \right) 
\,.
\end{equation}
Exploiting spherical symmetry, it is sufficient to analyze the Lagrangian above by considering $\theta=\pi/2$, which implies $\sin \theta = 1$. Utilizing the Euler-Lagrange equation $\frac{d}{d \lambda} \left( \frac{\partial \mathcal{L}}{\partial \dot{x}^\mu} \right) - \frac{\partial \mathcal{L}}{\partial x^\mu} = 0$ on the $t$ and $\phi$ components yields two constants of motion, denoted by $E$ and $L$ as $
 E \, = \, A(r) \,  \, \dot{t} \, $ and 
$ L  =  r^2    \dot{\phi} .
$

When deriving the orbit equation, it is convenient to employ the first integral of motion for photons, which is $g_{\mu \nu} \dot{x}^\mu \dot{x}^\nu  =  0$. This leads to the following orbit equation: 
\begin{equation} \label{eorb}
    \left( \dfrac{dr}{d\phi} \right) ^2 \, = \, \frac{r^2}{B(r)}   \left(  \frac{r^2}{A(r)}
\dfrac{E^2 }{L^2} - 1 \right) \, .
\end{equation}
Here, the constant $L/E$, denoted as the impact parameter $b$, is defined. The function associated with $b^{-1}$ is akin to the effective potential for null particles \cite{Perlick:2021aok} and is expressed as:
\begin{equation}
    h(r)^2 = \frac{r^2}{A(r)}.
\end{equation}
Thus, the radius of the photon-sphere $r_{\rm ph}$ can be determined by setting $h(r)' = 0$, leading to 
\begin{equation}
    A(r)'r^2 - 2A(r)r = 0,
\end{equation}
hence, by solving the above equation, we obtain this expression
\begin{equation}\label{quinticG}
    2 r^{7} -6 M r^{6}+5760\pi c_6 l_{\rm Pl}^4 M^{3}\vert_{r = r_{\rm ph}} = 0.
\end{equation}
The Abel-Ruffini theorem restricts finding an analytical solution to Eq. \eqref{quinticG} \cite{ruffini1813riflessioni}, as this is a polynomial of degree $7$. Assuming $c_6$ is of unity (as stated in Ref. \cite{Calmet:2021lny}), it can be found that the photon-sphere radius will still be $r_{\rm ph} = 3M$ if $M$ ranges from Planck mass black holes, up to the theoretical mass limit for black holes ($\sim 2.7 \times 10^{11} M_{\odot}$) \cite{King_2015}. It turns out that the only way that the $c_6$ influences the black hole geometry is when $M$ offsets the tiny value of the Planck length.

\section{The black hole shadow} \label{sec4}

Consider an observer (can also be a detector) at $(t_{\rm o},r_{\rm o},\theta_{\rm o} = \pi/2, \phi_{\rm o})$, the angular radius of the shadow is defined as:
\begin{equation}
    \tan(\alpha_{\text{sh}}) = \lim_{\Delta x \to 0}\frac{\Delta y}{\Delta x} = \left(\frac{r^2}{B(r)}\right)^{1/2} \frac{d\phi}{dr} \bigg|_{r=r_{\rm o}}.
\end{equation}
Utilizing Eq. \eqref{eorb} and noting that $h(r \rightarrow r_{\rm ph}) = b_{\text{crit}}$, we find:
\begin{equation}
    \sin^2 \alpha \, = \, \dfrac{b_\text{crit}^2}{h(r_{\rm o})^2}   \,
\end{equation}
Here, in our specific case where we already know that for both quantum and astrophysical black holes, $r_{\rm ph} = 3M$:
\begin{equation}
    b_\text{crit}^2 = \frac{19683 M^6}{640\pi c_6 l^{4}+729 M^{4}}.
\end{equation}
Notice how the critical impact parameter increases when the third-order curvature $c_6$ is negative, and how it decreases when $c_6$ is positive. Now, we can then determine the observer-dependent shadow radius using the expression:
\begin{equation}
    R_{\rm sh} = \sqrt{\frac{19683 M^6}{640\pi c_6 l^{4}+729 M^{4}} \left( 1-\frac{2 M}{r_{\rm o}}+\frac{640 \pi  c l_{\rm Pl}^4 M^{3}}{r_{\rm o}^{7}}\right)},
\end{equation}
where a certain restriction to the sign of $c_6$ can be gleaned. That is, $c_6$ must be negative. Furthermore, for $R_{\rm sh}$ not to become infinite, the maximum mass requirement should be
\begin{equation}
    M_{\rm max} = \frac{2\sqrt{3}\sqrt[4]{40 \pi c_6} l_{\rm Pl}}{9} \sim 4.315 l_{\rm Pl}.
\end{equation}
This is around $9.392 \times 10^{-8}$ kg. Furthermore, we can find the minimum mass of the quantum black hole by approximating the shadow radius to be $R_{\rm sh} \sim l_{\rm Pl}$, so that
\begin{equation}
    M_{\rm min} \sim 0.192 l_{\rm Pl},
\end{equation}
which should be around $4.179 \times 10^{-9}$ kg.

Interestingly, we can get an approximate equation when $r_{\rm o} >> M$:
\begin{equation}
    R_{\rm sh} = 3\sqrt{3} M - \frac{3\sqrt{3} M^2}{r_{\rm o}} + \frac{320\sqrt{3}\pi c_6 l_{\rm Pl}^4 (M-r_{\rm o})}{243 M^3 r_{\rm o}} + \mathcal{O}(l_{\rm P}^8).
\end{equation}
If the observer or detector is very close to the black hole ($r_{\rm o} << M$), we find
\begin{equation}
    R_{\rm sh} = \frac{2 \sqrt{30 \pi  c_6} l_{\rm Pl}^{2} \left(2916 M^{7}+2 M r_{\rm o}^{6}-r_{\rm o}^{7}\right)}{243 M^{9/2} r_{\rm o}^{7/2}} - \frac{3 \sqrt{30} r_{\rm o}^{5/2} \left(2M - r_{\rm o} \right)}{160 \sqrt{\pi c_6 M} l_{\rm Pl}^{2}} + \mathcal{O}(l_{\rm Pl}^6)
\end{equation}

\section{Deflection angle in weak field gravitational limits } \label{sec5}
In this section, we outline the calculation of the deflection angle of the Schwarzschild black hole with quantum gravitational correction at third-order curvature. With the metric given in Eq. \eqref{eq:gqs}, the Jacobi metric, accommodating time-like geodesics, is expressed as \cite{Li:2019vhp, Gibbons:2015qja}
\begin{equation}\label{metricEG}
    dl^2=g_{ij}dx^{i}dx^{j} =[E^2-\mu^2A(r)]\left(\frac{B(r)}{A(r)}dr^2+\frac{r^2}{A(r)}d\Omega^2\right),
\end{equation}
Here, $\mu$ denotes the particle's mass (which can be set to unity), and $E$ represents the energy of the massive particle, defined as:
\begin{equation} \label{en}
    E = \frac{1}{\sqrt{1-v^2}},
\end{equation}
where $v$ denotes the particle's velocity. Without loss of generality. we refer to the equatorial plane, where $\theta = \pi/2$. The Jacobi metric  simplifies to:
\begin{equation} \label{eJac}
    dl^2=(E^2-A(r))\left(\frac{B(r)}{A(r)}dr^2+\frac{r^2}{A(r)}d\phi^2\right)
\end{equation}
The determinant of the Jacobi metric above can also be easily computed as:
\begin{equation}
    g=\frac{B(r)r^2}{A(r)^2}(E^2-A(r))^2.
\end{equation}

Next, we will utilize these equations to determine the weak deflection angle using the Gauss-Bonnet theorem (GBT), which is originally stated as \cite{Gibbons:2008rj}:
\begin{equation} \label{eGBT}
    \iint_DKdS+\sum\limits_{a=1}^N \int_{\partial D_{a}} \kappa_{\text{g}} d\ell+ \sum\limits_{a=1}^N \theta_{a} = 2\pi\chi(D),
\end{equation}
Here, $D$ is a smooth, compact, freely orientable surface with boundary $\partial M$, embedded in $\mathbb{R}^3$. Let $K$ denote the Gaussian curvature of $D$ at each point. \textcolor{black}{$\chi(D)$ represents the Euler characteristic of $D$ (equal to $1$ since $D$ is in a non-singular region)}, $\int\int_D K \, dA$ denotes the integral of the Gaussian curvature over the surface $D$, $\int_{\partial D_a} k_g \, d\ell$  represents the integral of the geodesic curvature along the boundary $\partial D$. In this context, $k_g$ denotes the geodesic curvature along the boundary. Additionally, $\theta_{a}$ is the jump angle.

It has been demonstrated by \cite{Li:2020wvn} that in a spherically symmetric static (SSS) spacetime without asymptotic flatness, Eq. \eqref{eGBT} can be expressed as
\begin{equation} \label{deflection}
    \hat{\alpha} = \iint_{D}KdS + \phi_{\text{RS}},
\end{equation}
Here, $\text{S}$ and $\text{R}$ represent the radial positions of the source and receiver, respectively (see for details \cite{Ishihara:2016vdc,Ishihara:2016sfv,Li:2020wvn}), and $\phi_{\rm RS} = \phi_{\rm R} - \phi_{\rm S}$ is the positional angle along the equatorial plane. These serve as the integration domains, and it's worth noting that the infinitesimal curved surface $dS$ is defined by:
\begin{equation}
    dS = \sqrt{g}drd\phi.
\end{equation}
Additionally, $\phi_{\rm RS}$ represents the coordinate position angle between the source and the receiver, defined as $\phi_{\rm RS} = \phi_{\rm R}-\phi_{\rm S}$, which can be determined through the iterative solution of
\begin{equation}
    F(u) = \left(\frac{du}{d\phi}\right)^2  = \frac{1}{A(u)B(u)}\Bigg[\left(\frac{E}{J}\right)^2-A(u)\left(\frac{1}{J^2}+u^2\right)\Bigg].
\end{equation}
Here, we have utilized the substitution $r = 1/u$ and the angular momentum of the massive particle given by:
\begin{equation}
    J = v b E,
\end{equation}
where $b$ represents the impact parameter.

Given the metric coefficients, we obtain the orbit equation in terms of $u$ as:
\begin{equation} \label{e_time-orb}
    F(u) = \frac{E^2-1}{J^2} -u^{2}+2 u \left(u^{2}+\frac{1}{J^2} \right) M +3456 \pi  c_6 l_{\rm Pl}^4 \,u^{6} \left(-u^{2}+\frac{E^2-1}{J^2} \right) M^{2}.
\end{equation}
For a timelike orbit, an excellent approximate solution for the Schwarzschild case is the expression:
\begin{equation} \label{euphi}
    u(\phi) = \frac{\sin(\phi)}{b}+\frac{1+v^2\cos^2(\phi)}{b^2v^2}M.
\end{equation}
Instead of attempting to solve the challenging differential equation presented in Eq. \eqref{e_time-orb}, which incorporates the quantum correction parameter, we can employ iterative methods by introducing a term to Eq. \eqref{euphi} with $p c_6 l_{\rm Pl}^4 M^2$
\begin{equation} \label{euphi2}
    u(\phi) = \frac{\sin(\phi)}{b}+\frac{1+v^2\cos^2(\phi)}{b^2v^2}M + p c_6 l_{\rm Pl}^4 M^2,
\end{equation}
where $p$ is some unknown coefficient coupled to the parameter $c_6$, which is later to be determined. The objective is to determine $p$ through iteration. Upon solving for $p$, we found $p=0$, indicating that the quantum correction appears not to affect $u(\phi)$. For future reference, we solved for $\phi$ as:
\begin{equation} \label{eqphi}
    \phi = \arcsin(bu)+\frac{M\left[v^{2}\left(b^{2}u^{2}-1\right)-1\right]}{bv^{2}\sqrt{1-b^{2}u^{2}}}.
\end{equation}

The Gaussian curvature $K$, expressed in terms of affine connections and the determinant $g$, is defined as: 
\begin{align}
    K=\frac{1}{\sqrt{g}}\left[\frac{\partial}{\partial\phi}\left(\frac{\sqrt{g}}{g_{rr}}\Gamma_{rr}^{\phi}\right)-\frac{\partial}{\partial r}\left(\frac{\sqrt{g}}{g_{rr}}\Gamma_{r\phi}^{\phi}\right)\right] 
    =-\frac{1}{\sqrt{g}}\left[\frac{\partial}{\partial r}\left(\frac{\sqrt{g}}{g_{rr}}\Gamma_{r\phi}^{\phi}\right)\right]
\end{align}
since $\Gamma_{rr}^{\phi} = 0$ for Eq. \eqref{eJac}. Then with the analytical solution to $r_{\rm ph}$,
\begin{equation}
    \left[\int K\sqrt{g}dr\right]\bigg|_{r=r_{\rm ph}} = 0,
\end{equation}
thus,
\begin{equation} \label{gct}
    \int_{r_{\rm ph}}^{r(\phi)} K\sqrt{g}dr = -\frac{2r A(r)\left(E^{2}-A(r)\right)-E^{2}r^2A(r)'}{2r A(r)\left(E^{2}-A(r)\right)\sqrt{B(r)}}\bigg|_{r = r(\phi)}.
\end{equation}
The prime notation indicates differentiation with respect to $r$. The weak deflection angle is then given by \cite{Li:2020wvn}:
\begin{align} \label{eqwda}
    \hat{\alpha} = \int^{\phi_{\rm R}}_{\phi_{\rm S}} \left[-\frac{2r A(r)\left(E^{2}-A(r)\right)-E^{2}r^2A(r)'}{2rA(r)\left(E^{2}-A(r)\right)\sqrt{B(r)}}\bigg|_{r = r(\phi)}\right] d\phi + \phi_{\rm RS}.
\end{align}
Using Eq. \eqref{metricEG} in Eq. \eqref{gct}, we find
\begin{align} \label{gct2}
    &\left[\int K\sqrt{g}dr\right]\bigg|_{r=r_\phi} = -\frac{\left(2E^{2}-1\right)M(\cos(\phi_{\rm R})-\cos(\phi_{\rm S}))}{\left(E^{2}-1\right)b} -\phi_{\rm RS} 
    -\frac{108 \pi c_6 l_{\rm Pl}^4 M^{2}}{b^{6}} \times \\
    & 
    \times \left[ \left(-\frac{8 (\cos^6(\phi_{\rm R})-\cos^5(\phi_{\rm S}))}{3}+\frac{26 (\cos^3(\phi_{\rm R})-\cos^3(\phi_{\rm S}))}{3}-11 (\cos(\phi_{\rm R})-\cos(\phi_{\rm S}))\right) (\sin(\phi_{\rm R})-\sin(\phi_{\rm S}))+\phi_{\rm RS}\right]. \nonumber
\end{align}
From Eq. \eqref{eqphi}, we obtain the equatorial angles for the source and receiver:
\begin{equation} \label{s}
    \phi_{\rm S} =\arcsin(bu)+\frac{M\left[v^{2}\left(b^{2}u^{2}-1\right)-1\right]}{bv^{2}\sqrt{1-b^{2}u^{2}}},
\end{equation}
\begin{equation} \label{r}
    \phi_{\rm R} =\pi -\arcsin(bu)-\frac{M\left[v^{2}\left(b^{2}u^{2}-1\right)-1\right]}{bv^{2}\sqrt{1-b^{2}u^{2}}},
\end{equation}
respectively. Upon careful observation of these equations, we can express $\phi_{\rm RS} = \pi - 2\phi_{\rm S}$. Now, let's take note of the following relations:
\begin{equation}
    \cos(\pi-\phi_{\rm S})=-\cos(\phi_{\rm S}), \qquad
    \sin(\pi-\phi_{\rm S})=\sin(\phi_{\rm S}).
\end{equation}
The last property cancels the sine terms in Eq. \eqref{gct2}. We find $\cos(\phi_{\rm S})$ as
\begin{equation} \label{cs}
    \cos(\phi_{\rm S}) = \sqrt{1-b^{2}u_{\rm S}^{2}}-\frac{Mu_{\rm S}\left[v^{2}\left(b^{2}u_{\rm S}^{2}-1\right)-1\right]}{v^{2}\sqrt{\left(1-b^{2}u_{\rm S}^{2}\right)}}.
\end{equation}
For $\cos^3(\phi_{\rm S})$ and $\cos^5(\phi_{\rm S})$, these are
\begin{equation}
    \cos^3(\phi_{\rm S}) = \left(-b^{2} u_{\rm S}^{2}+1\right)^{\frac{3}{2}}-\frac{3 \sqrt{-b^{2} u_{\rm S}^{2}+1}\, u_{\rm S} \left(-1+\left(b^{2} u_{\rm S}^{2}-1\right) v^{2}\right) M}{2 v^{2}},
\end{equation}
and
\begin{equation}
    \cos^5(\phi_{\rm S}) = \left(-b^{2} u_{\rm S}^{2}+1\right)^{\frac{5}{2}}-\frac{5 \left(-b^{2} u_{\rm S}^{2}+1\right)^{\frac{3}{2}} u_{\rm S} \left(b^{2} u_{\rm S}^{2} v^{2}-v^{2}-1\right) M}{2 v^{2}},
\end{equation}
respectively. Finally, for $\sin(\phi_{\rm S})$,
\begin{equation}  \label{488}
    \sin (\phi_{\rm S}) = b u_{\rm S} +\frac{\left[-1+\left(b^{2} u_{\rm S}^{2}-1\right) v^{2}\right] M}{2 b \,v^{2}}.
\end{equation}
Using Eq. \eqref{en} and by plugging Eqs. \eqref{s}-\eqref{488} into Eq. \eqref{eqwda}, we finally obtain
\begin{align}
    \alpha &= \frac{\left(v^{2}+1\right) M}{b \,v^{2}}\left( \sqrt{1-b^{2} u_{\rm S}^{2}}+\sqrt{1-b^{2} u_{\rm R}^{2}} \right) \nonumber \\
    &- \frac{270 \pi  c_6 l_{\rm Pl}^4 M^{2}}{b^{6}} \left[ \arcsin \! \left(b u_{\rm S} \right)+ 2\, b u_{\rm S} \sqrt{-b^{2} u_{\rm S}^{2}+1} +\arcsin \! \left(b u_{\rm R} \right) + 2\, b u_{\rm R} \sqrt{-b^{2} u_{\rm R}^{2}+1} \right].
\end{align}

The above expression is general, considering the finite distance of both the source and the receiver from the black hole. However, when both $u_{\rm S}$ and $u_{\rm R}$ approach zero, the above expressions simplify to:
\begin{equation}
    \alpha = \frac{2\left(v^{2}+1\right) M}{b \,v^{2}}.
\end{equation}
The above expressions can be readily observed to reduce to $\alpha = 4M/b$ for photons ($v=1$).

This outcome is remarkable since the quantum correction parameter $c_6$ only manifests when considering the non-zero value for $u$. It implies that when an observer is sufficiently distant from the black hole, the influence of $c_6$ diminishes. Such an effect is the same with the shadow behavior when $r_{\rm o}>>M$. In addition, the term with the quantum correction does not involve the speed of the time-like particle. The expression also ensures that for quantum correction to manifest for this deflection angle, the impact parameter should be very small and compensated with increased $u$ (or $r \sim 0$).

\section{Conclusion} \label{conc}
This study investigated the impact of the quantum-gravity correction parameter $c_6$ at the third-order curvature on black hole shadows and weak deflection angles. These are achieved by studying the photon-sphere and employing the finite-distance version of the GBT.

Despite the challenges posed by the analytical intractability of the resulting expression in Eq. \eqref{quinticG}, the result suggested that $r_{\rm ph} = 3M$ for both quantum and astrophysical black holes, due to the influence of the Planck's length. The information allowed us to derive analytically the observer-dependent shadow radius. The exact formula revealed some information about the minimum and maximum mass a black hole can have due to the third-order curvature correction. That is $0.192 \lesssim M/l_{\rm Pl} \lesssim 4.315$. Also, the derived analytic formula for the near and far approximation of $R_{\rm sh}$ implied that $c_6$'s effect would only be valid for quantum black holes and if $r_{\rm o}$ near the Planck length.

Finally, the examination of weak deflection angle behavior suggests that the deviation would only occur if $M$ and $b$ are around the order of $l_{\rm Pl}$, still for quantum black holes. Furthermore, a special condition also exists: for this correction to occur, finite distance expressed inversely as $u$ must be non-zero. It also turned out that the correction does not involve the deflection of time-like particles.

As this theoretical analysis provided valuable insights into where quantum black holes are involved, detecting the deviations from the classical Schwarzschild case caused by the third-order curvature will remain challenging due to the difficulty of probing length scales comparable to the Planck length.

\begin{acknowledgements}
A. {\"O}., G.L., and R. P. would like to acknowledge networking support of the COST Action CA18108 - Quantum gravity phenomenology in the multi-messenger approach (QG-MM), COST Action CA21106 - COSMIC WISPers in the Dark Universe: Theory, astrophysics and experiments (CosmicWISPers), the COST Action CA22113 - Fundamental challenges in theoretical physics (THEORY-CHALLENGES), and the COST Action CA21136 - Addressing observational tensions in cosmology with systematics and fundamental physics (CosmoVerse). We also thank TUBITAK and SCOAP3 for their support.
\end{acknowledgements}

\bibliography{ref}

\begin{thebibliography}{82}
\expandafter\ifx\csname natexlab\endcsname\relax\def\natexlab#1{#1}\fi
\expandafter\ifx\csname bibnamefont\endcsname\relax
  \def\bibnamefont#1{#1}\fi
\expandafter\ifx\csname bibfnamefont\endcsname\relax
  \def\bibfnamefont#1{#1}\fi
\expandafter\ifx\csname citenamefont\endcsname\relax
  \def\citenamefont#1{#1}\fi
\expandafter\ifx\csname url\endcsname\relax
  \def\url#1{\texttt{#1}}\fi
\expandafter\ifx\csname urlprefix\endcsname\relax\def\urlprefix{URL }\fi
\providecommand{\bibinfo}[2]{#2}
\providecommand{\eprint}[2][]{\url{#2}}

\bibitem[{\citenamefont{Calmet and El-Menoufi}(2017)}]{Calmet:2017qqa}
\bibinfo{author}{\bibfnamefont{X.}~\bibnamefont{Calmet}} \bibnamefont{and} \bibinfo{author}{\bibfnamefont{B.~K.} \bibnamefont{El-Menoufi}}, \bibinfo{journal}{Eur. Phys. J. C} \textbf{\bibinfo{volume}{77}}, \bibinfo{pages}{243} (\bibinfo{year}{2017}), \eprint{1704.00261}.

\bibitem[{\citenamefont{Calmet}(2018)}]{Calmet:2018elv}
\bibinfo{author}{\bibfnamefont{X.}~\bibnamefont{Calmet}}, \bibinfo{journal}{Phys. Lett. B} \textbf{\bibinfo{volume}{787}}, \bibinfo{pages}{36} (\bibinfo{year}{2018}), \eprint{1810.09719}.

\bibitem[{\citenamefont{Calmet and Kuipers}(2021)}]{Calmet:2021lny}
\bibinfo{author}{\bibfnamefont{X.}~\bibnamefont{Calmet}} \bibnamefont{and} \bibinfo{author}{\bibfnamefont{F.}~\bibnamefont{Kuipers}}, \bibinfo{journal}{Phys. Rev. D} \textbf{\bibinfo{volume}{104}}, \bibinfo{pages}{066012} (\bibinfo{year}{2021}), \eprint{2108.06824}.

\bibitem[{\citenamefont{Calmet et~al.}(2023)\citenamefont{Calmet, Hsu, and Sebastianutti}}]{Calmet:2023gbw}
\bibinfo{author}{\bibfnamefont{X.}~\bibnamefont{Calmet}}, \bibinfo{author}{\bibfnamefont{S.~D.~H.} \bibnamefont{Hsu}}, \bibnamefont{and} \bibinfo{author}{\bibfnamefont{M.}~\bibnamefont{Sebastianutti}}, \bibinfo{journal}{Phys. Lett. B} \textbf{\bibinfo{volume}{841}}, \bibinfo{pages}{137820} (\bibinfo{year}{2023}), \eprint{2303.00310}.

\bibitem[{\citenamefont{Kiefer}(2020)}]{Kiefer:2020cbu}
\bibinfo{author}{\bibfnamefont{C.}~\bibnamefont{Kiefer}}, \bibinfo{journal}{J. Phys. Conf. Ser.} \textbf{\bibinfo{volume}{1612}}, \bibinfo{pages}{012017} (\bibinfo{year}{2020}).

\bibitem[{\citenamefont{Binetti et~al.}(2022)\citenamefont{Binetti, Del~Piano, Hohenegger, Pezzella, and Sannino}}]{Binetti:2022xdi}
\bibinfo{author}{\bibfnamefont{E.}~\bibnamefont{Binetti}}, \bibinfo{author}{\bibfnamefont{M.}~\bibnamefont{Del~Piano}}, \bibinfo{author}{\bibfnamefont{S.}~\bibnamefont{Hohenegger}}, \bibinfo{author}{\bibfnamefont{F.}~\bibnamefont{Pezzella}}, \bibnamefont{and} \bibinfo{author}{\bibfnamefont{F.}~\bibnamefont{Sannino}}, \bibinfo{journal}{Phys. Rev. D} \textbf{\bibinfo{volume}{106}}, \bibinfo{pages}{046006} (\bibinfo{year}{2022}), \eprint{2203.13515}.

\bibitem[{\citenamefont{Mertens and Turiaci}(2023)}]{Mertens:2022irh}
\bibinfo{author}{\bibfnamefont{T.~G.} \bibnamefont{Mertens}} \bibnamefont{and} \bibinfo{author}{\bibfnamefont{G.~J.} \bibnamefont{Turiaci}}, \bibinfo{journal}{Living Rev. Rel.} \textbf{\bibinfo{volume}{26}}, \bibinfo{pages}{4} (\bibinfo{year}{2023}), \eprint{2210.10846}.

\bibitem[{\citenamefont{Del~Piano et~al.}(2024)\citenamefont{Del~Piano, Hohenegger, and Sannino}}]{DelPiano:2023fiw}
\bibinfo{author}{\bibfnamefont{M.}~\bibnamefont{Del~Piano}}, \bibinfo{author}{\bibfnamefont{S.}~\bibnamefont{Hohenegger}}, \bibnamefont{and} \bibinfo{author}{\bibfnamefont{F.}~\bibnamefont{Sannino}}, \bibinfo{journal}{Phys. Rev. D} \textbf{\bibinfo{volume}{109}}, \bibinfo{pages}{024045} (\bibinfo{year}{2024}), \eprint{2307.13489}.

\bibitem[{\citenamefont{Devi et~al.}(2023)\citenamefont{Devi, S., Chakrabarti, and Majhi}}]{Devi:2021ctm}
\bibinfo{author}{\bibfnamefont{S.}~\bibnamefont{Devi}}, \bibinfo{author}{\bibfnamefont{A.~N.} \bibnamefont{S.}}, \bibinfo{author}{\bibfnamefont{S.}~\bibnamefont{Chakrabarti}}, \bibnamefont{and} \bibinfo{author}{\bibfnamefont{B.~R.} \bibnamefont{Majhi}}, \bibinfo{journal}{Phys. Dark Univ.} \textbf{\bibinfo{volume}{39}}, \bibinfo{pages}{101173} (\bibinfo{year}{2023}), \eprint{2105.11847}.

\bibitem[{\citenamefont{Saadati and Shojai}(2024)}]{Saadati:2023jym}
\bibinfo{author}{\bibfnamefont{R.}~\bibnamefont{Saadati}} \bibnamefont{and} \bibinfo{author}{\bibfnamefont{F.}~\bibnamefont{Shojai}}, \bibinfo{journal}{Class. Quant. Grav.} \textbf{\bibinfo{volume}{41}}, \bibinfo{pages}{015032} (\bibinfo{year}{2024}), \eprint{2312.12087}.

\bibitem[{\citenamefont{Jiang et~al.}(2024)\citenamefont{Jiang, Liu, Dihingia, Mizuno, Xu, Zhu, and Wu}}]{Jiang:2023img}
\bibinfo{author}{\bibfnamefont{H.-X.} \bibnamefont{Jiang}}, \bibinfo{author}{\bibfnamefont{C.}~\bibnamefont{Liu}}, \bibinfo{author}{\bibfnamefont{I.~K.} \bibnamefont{Dihingia}}, \bibinfo{author}{\bibfnamefont{Y.}~\bibnamefont{Mizuno}}, \bibinfo{author}{\bibfnamefont{H.}~\bibnamefont{Xu}}, \bibinfo{author}{\bibfnamefont{T.}~\bibnamefont{Zhu}}, \bibnamefont{and} \bibinfo{author}{\bibfnamefont{Q.}~\bibnamefont{Wu}}, \bibinfo{journal}{JCAP} \textbf{\bibinfo{volume}{01}}, \bibinfo{pages}{059} (\bibinfo{year}{2024}), \eprint{2312.04288}.

\bibitem[{\citenamefont{Lambiase et~al.}(2023)\citenamefont{Lambiase, Pantig, Gogoi, and \"Ovg\"un}}]{Lambiase:2023hng}
\bibinfo{author}{\bibfnamefont{G.}~\bibnamefont{Lambiase}}, \bibinfo{author}{\bibfnamefont{R.~C.} \bibnamefont{Pantig}}, \bibinfo{author}{\bibfnamefont{D.~J.} \bibnamefont{Gogoi}}, \bibnamefont{and} \bibinfo{author}{\bibfnamefont{A.}~\bibnamefont{\"Ovg\"un}}, \bibinfo{journal}{Eur. Phys. J. C} \textbf{\bibinfo{volume}{83}}, \bibinfo{pages}{679} (\bibinfo{year}{2023}), \eprint{2304.00183}.

\bibitem[{\citenamefont{Atamurotov et~al.}(2023)\citenamefont{Atamurotov, Jamil, and Jusufi}}]{Atamurotov:2022iwj}
\bibinfo{author}{\bibfnamefont{F.}~\bibnamefont{Atamurotov}}, \bibinfo{author}{\bibfnamefont{M.}~\bibnamefont{Jamil}}, \bibnamefont{and} \bibinfo{author}{\bibfnamefont{K.}~\bibnamefont{Jusufi}}, \bibinfo{journal}{Chin. Phys. C} \textbf{\bibinfo{volume}{47}}, \bibinfo{pages}{035106} (\bibinfo{year}{2023}), \eprint{2212.12949}.

\bibitem[{\citenamefont{Yang et~al.}(2023)\citenamefont{Yang, Zhang, and Ma}}]{Yang:2022btw}
\bibinfo{author}{\bibfnamefont{J.}~\bibnamefont{Yang}}, \bibinfo{author}{\bibfnamefont{C.}~\bibnamefont{Zhang}}, \bibnamefont{and} \bibinfo{author}{\bibfnamefont{Y.}~\bibnamefont{Ma}}, \bibinfo{journal}{Eur. Phys. J. C} \textbf{\bibinfo{volume}{83}}, \bibinfo{pages}{619} (\bibinfo{year}{2023}), \eprint{2211.04263}.

\bibitem[{\citenamefont{Tang and Xu}(2022)}]{Tang:2022uwi}
\bibinfo{author}{\bibfnamefont{M.}~\bibnamefont{Tang}} \bibnamefont{and} \bibinfo{author}{\bibfnamefont{Z.}~\bibnamefont{Xu}}, \bibinfo{journal}{JHEP} \textbf{\bibinfo{volume}{12}}, \bibinfo{pages}{125} (\bibinfo{year}{2022}), \eprint{2209.08202}.

\bibitem[{\citenamefont{Lobos and Pantig}(2022)}]{Lobos:2022jsz}
\bibinfo{author}{\bibfnamefont{N.~J. L.~S.} \bibnamefont{Lobos}} \bibnamefont{and} \bibinfo{author}{\bibfnamefont{R.~C.} \bibnamefont{Pantig}}, \bibinfo{journal}{MDPI Physics} \textbf{\bibinfo{volume}{4}}, \bibinfo{pages}{1318} (\bibinfo{year}{2022}), \eprint{2208.00618}.

\bibitem[{\citenamefont{Xu and Tang}(2022)}]{Xu:2021xgw}
\bibinfo{author}{\bibfnamefont{Z.}~\bibnamefont{Xu}} \bibnamefont{and} \bibinfo{author}{\bibfnamefont{M.}~\bibnamefont{Tang}}, \bibinfo{journal}{Chin. Phys. C} \textbf{\bibinfo{volume}{46}}, \bibinfo{pages}{085101} (\bibinfo{year}{2022}), \eprint{2109.14245}.

\bibitem[{\citenamefont{Zakharov}(2022)}]{Zakharov:2021gbg}
\bibinfo{author}{\bibfnamefont{A.~F.} \bibnamefont{Zakharov}}, \bibinfo{journal}{Universe} \textbf{\bibinfo{volume}{8}}, \bibinfo{pages}{141} (\bibinfo{year}{2022}), \eprint{2108.01533}.

\bibitem[{\citenamefont{Zakharov}(2014)}]{Zakharov:2014lqa}
\bibinfo{author}{\bibfnamefont{A.~F.} \bibnamefont{Zakharov}}, \bibinfo{journal}{Phys. Rev. D} \textbf{\bibinfo{volume}{90}}, \bibinfo{pages}{062007} (\bibinfo{year}{2014}), \eprint{1407.7457}.

\bibitem[{\citenamefont{Vagnozzi et~al.}(2023)}]{Vagnozzi:2022moj}
\bibinfo{author}{\bibfnamefont{S.}~\bibnamefont{Vagnozzi}} \bibnamefont{et~al.}, \bibinfo{journal}{Class. Quant. Grav.} \textbf{\bibinfo{volume}{40}}, \bibinfo{pages}{165007} (\bibinfo{year}{2023}), \eprint{2205.07787}.

\bibitem[{\citenamefont{Vagnozzi and Visinelli}(2019)}]{Vagnozzi:2019apd}
\bibinfo{author}{\bibfnamefont{S.}~\bibnamefont{Vagnozzi}} \bibnamefont{and} \bibinfo{author}{\bibfnamefont{L.}~\bibnamefont{Visinelli}}, \bibinfo{journal}{Phys. Rev. D} \textbf{\bibinfo{volume}{100}}, \bibinfo{pages}{024020} (\bibinfo{year}{2019}), \eprint{1905.12421}.

\bibitem[{\citenamefont{Allahyari et~al.}(2020)\citenamefont{Allahyari, Khodadi, Vagnozzi, and Mota}}]{Allahyari:2019jqz}
\bibinfo{author}{\bibfnamefont{A.}~\bibnamefont{Allahyari}}, \bibinfo{author}{\bibfnamefont{M.}~\bibnamefont{Khodadi}}, \bibinfo{author}{\bibfnamefont{S.}~\bibnamefont{Vagnozzi}}, \bibnamefont{and} \bibinfo{author}{\bibfnamefont{D.~F.} \bibnamefont{Mota}}, \bibinfo{journal}{JCAP} \textbf{\bibinfo{volume}{02}}, \bibinfo{pages}{003} (\bibinfo{year}{2020}), \eprint{1912.08231}.

\bibitem[{\citenamefont{Khodadi et~al.}(2020)\citenamefont{Khodadi, Allahyari, Vagnozzi, and Mota}}]{Khodadi:2020jij}
\bibinfo{author}{\bibfnamefont{M.}~\bibnamefont{Khodadi}}, \bibinfo{author}{\bibfnamefont{A.}~\bibnamefont{Allahyari}}, \bibinfo{author}{\bibfnamefont{S.}~\bibnamefont{Vagnozzi}}, \bibnamefont{and} \bibinfo{author}{\bibfnamefont{D.~F.} \bibnamefont{Mota}}, \bibinfo{journal}{JCAP} \textbf{\bibinfo{volume}{09}}, \bibinfo{pages}{026} (\bibinfo{year}{2020}), \eprint{2005.05992}.

\bibitem[{\citenamefont{Meng et~al.}(2023)\citenamefont{Meng, Kuang, Wang, Wang, and Wu}}]{Meng:2023htc}
\bibinfo{author}{\bibfnamefont{Y.}~\bibnamefont{Meng}}, \bibinfo{author}{\bibfnamefont{X.-M.} \bibnamefont{Kuang}}, \bibinfo{author}{\bibfnamefont{X.-J.} \bibnamefont{Wang}}, \bibinfo{author}{\bibfnamefont{B.}~\bibnamefont{Wang}}, \bibnamefont{and} \bibinfo{author}{\bibfnamefont{J.-P.} \bibnamefont{Wu}}, \bibinfo{journal}{Phys. Rev. D} \textbf{\bibinfo{volume}{108}}, \bibinfo{pages}{064013} (\bibinfo{year}{2023}), \eprint{2306.10459}.

\bibitem[{\citenamefont{Li and Kuang}(2023)}]{Li:2023djs}
\bibinfo{author}{\bibfnamefont{Y.-Z.} \bibnamefont{Li}} \bibnamefont{and} \bibinfo{author}{\bibfnamefont{X.-M.} \bibnamefont{Kuang}}, \bibinfo{journal}{Phys. Rev. D} \textbf{\bibinfo{volume}{107}}, \bibinfo{pages}{064052} (\bibinfo{year}{2023}).

\bibitem[{\citenamefont{Kuang et~al.}(2022)\citenamefont{Kuang, Tang, Wang, and Wang}}]{Kuang:2022ojj}
\bibinfo{author}{\bibfnamefont{X.-M.} \bibnamefont{Kuang}}, \bibinfo{author}{\bibfnamefont{Z.-Y.} \bibnamefont{Tang}}, \bibinfo{author}{\bibfnamefont{B.}~\bibnamefont{Wang}}, \bibnamefont{and} \bibinfo{author}{\bibfnamefont{A.}~\bibnamefont{Wang}}, \bibinfo{journal}{Phys. Rev. D} \textbf{\bibinfo{volume}{106}}, \bibinfo{pages}{064012} (\bibinfo{year}{2022}), \eprint{2206.05878}.

\bibitem[{\citenamefont{Abdujabbarov et~al.}(2016)\citenamefont{Abdujabbarov, Amir, Ahmedov, and Ghosh}}]{Abdujabbarov:2016hnw}
\bibinfo{author}{\bibfnamefont{A.}~\bibnamefont{Abdujabbarov}}, \bibinfo{author}{\bibfnamefont{M.}~\bibnamefont{Amir}}, \bibinfo{author}{\bibfnamefont{B.}~\bibnamefont{Ahmedov}}, \bibnamefont{and} \bibinfo{author}{\bibfnamefont{S.~G.} \bibnamefont{Ghosh}}, \bibinfo{journal}{Phys. Rev. D} \textbf{\bibinfo{volume}{93}}, \bibinfo{pages}{104004} (\bibinfo{year}{2016}), \eprint{1604.03809}.

\bibitem[{\citenamefont{Atamurotov et~al.}(2013)\citenamefont{Atamurotov, Abdujabbarov, and Ahmedov}}]{Atamurotov:2013sca}
\bibinfo{author}{\bibfnamefont{F.}~\bibnamefont{Atamurotov}}, \bibinfo{author}{\bibfnamefont{A.}~\bibnamefont{Abdujabbarov}}, \bibnamefont{and} \bibinfo{author}{\bibfnamefont{B.}~\bibnamefont{Ahmedov}}, \bibinfo{journal}{Phys. Rev. D} \textbf{\bibinfo{volume}{88}}, \bibinfo{pages}{064004} (\bibinfo{year}{2013}).

\bibitem[{\citenamefont{Atamurotov et~al.}(2016)\citenamefont{Atamurotov, Ghosh, and Ahmedov}}]{Atamurotov:2015xfa}
\bibinfo{author}{\bibfnamefont{F.}~\bibnamefont{Atamurotov}}, \bibinfo{author}{\bibfnamefont{S.~G.} \bibnamefont{Ghosh}}, \bibnamefont{and} \bibinfo{author}{\bibfnamefont{B.}~\bibnamefont{Ahmedov}}, \bibinfo{journal}{Eur. Phys. J. C} \textbf{\bibinfo{volume}{76}}, \bibinfo{pages}{273} (\bibinfo{year}{2016}), \eprint{1506.03690}.

\bibitem[{\citenamefont{Islam et~al.}(2020)\citenamefont{Islam, Kumar, and Ghosh}}]{Islam:2020xmy}
\bibinfo{author}{\bibfnamefont{S.~U.} \bibnamefont{Islam}}, \bibinfo{author}{\bibfnamefont{R.}~\bibnamefont{Kumar}}, \bibnamefont{and} \bibinfo{author}{\bibfnamefont{S.~G.} \bibnamefont{Ghosh}}, \bibinfo{journal}{JCAP} \textbf{\bibinfo{volume}{09}}, \bibinfo{pages}{030} (\bibinfo{year}{2020}), \eprint{2004.01038}.

\bibitem[{\citenamefont{Kumar and Ghosh}(2020)}]{Kumar:2020owy}
\bibinfo{author}{\bibfnamefont{R.}~\bibnamefont{Kumar}} \bibnamefont{and} \bibinfo{author}{\bibfnamefont{S.~G.} \bibnamefont{Ghosh}}, \bibinfo{journal}{JCAP} \textbf{\bibinfo{volume}{07}}, \bibinfo{pages}{053} (\bibinfo{year}{2020}), \eprint{2003.08927}.

\bibitem[{\citenamefont{Afrin et~al.}(2021)\citenamefont{Afrin, Kumar, and Ghosh}}]{Afrin:2021imp}
\bibinfo{author}{\bibfnamefont{M.}~\bibnamefont{Afrin}}, \bibinfo{author}{\bibfnamefont{R.}~\bibnamefont{Kumar}}, \bibnamefont{and} \bibinfo{author}{\bibfnamefont{S.~G.} \bibnamefont{Ghosh}}, \bibinfo{journal}{Mon. Not. Roy. Astron. Soc.} \textbf{\bibinfo{volume}{504}}, \bibinfo{pages}{5927} (\bibinfo{year}{2021}), \eprint{2103.11417}.

\bibitem[{\citenamefont{Tsukamoto et~al.}(2014)\citenamefont{Tsukamoto, Li, and Bambi}}]{Tsukamoto:2014tja}
\bibinfo{author}{\bibfnamefont{N.}~\bibnamefont{Tsukamoto}}, \bibinfo{author}{\bibfnamefont{Z.}~\bibnamefont{Li}}, \bibnamefont{and} \bibinfo{author}{\bibfnamefont{C.}~\bibnamefont{Bambi}}, \bibinfo{journal}{JCAP} \textbf{\bibinfo{volume}{06}}, \bibinfo{pages}{043} (\bibinfo{year}{2014}), \eprint{1403.0371}.

\bibitem[{\citenamefont{Tsukamoto}(2018)}]{Tsukamoto:2017fxq}
\bibinfo{author}{\bibfnamefont{N.}~\bibnamefont{Tsukamoto}}, \bibinfo{journal}{Phys. Rev. D} \textbf{\bibinfo{volume}{97}}, \bibinfo{pages}{064021} (\bibinfo{year}{2018}), \eprint{1708.07427}.

\bibitem[{\citenamefont{\"Ovg\"un et~al.}(2023{\natexlab{a}})\citenamefont{\"Ovg\"un, Sese, and Pantig}}]{Ovgun:2023wmc}
\bibinfo{author}{\bibfnamefont{A.}~\bibnamefont{\"Ovg\"un}}, \bibinfo{author}{\bibfnamefont{L.~J.~F.} \bibnamefont{Sese}}, \bibnamefont{and} \bibinfo{author}{\bibfnamefont{R.~C.} \bibnamefont{Pantig}}, \bibinfo{journal}{Annalen Phys.} \textbf{\bibinfo{volume}{2023}}, \bibinfo{pages}{2300390} (\bibinfo{year}{2023}{\natexlab{a}}), \eprint{2309.07442}.

\bibitem[{\citenamefont{{Cunningham} and {Bardeen}}(1973)}]{Cunningham}
\bibinfo{author}{\bibfnamefont{C.~T.} \bibnamefont{{Cunningham}}} \bibnamefont{and} \bibinfo{author}{\bibfnamefont{J.~M.} \bibnamefont{{Bardeen}}}, \bibinfo{journal}{\apj} \textbf{\bibinfo{volume}{183}}, \bibinfo{pages}{237} (\bibinfo{year}{1973}).

\bibitem[{\citenamefont{{Bardeen}}(1974)}]{1974IAUS...64..132B}
\bibinfo{author}{\bibfnamefont{J.~M.} \bibnamefont{{Bardeen}}}, in \emph{\bibinfo{booktitle}{Gravitational Radiation and Gravitational Collapse}}, edited by \bibinfo{editor}{\bibfnamefont{C.}~\bibnamefont{{Dewitt-Morette}}} (\bibinfo{year}{1974}), vol.~\bibinfo{volume}{64}, p. \bibinfo{pages}{132}.

\bibitem[{\citenamefont{Luminet}(1979)}]{Luminet:1979nyg}
\bibinfo{author}{\bibfnamefont{J.~P.} \bibnamefont{Luminet}}, \bibinfo{journal}{Astron. Astrophys.} \textbf{\bibinfo{volume}{75}}, \bibinfo{pages}{228} (\bibinfo{year}{1979}).

\bibitem[{\citenamefont{Falcke et~al.}(2000)\citenamefont{Falcke, Melia, and Agol}}]{Falcke:1999pj}
\bibinfo{author}{\bibfnamefont{H.}~\bibnamefont{Falcke}}, \bibinfo{author}{\bibfnamefont{F.}~\bibnamefont{Melia}}, \bibnamefont{and} \bibinfo{author}{\bibfnamefont{E.}~\bibnamefont{Agol}}, \bibinfo{journal}{Astrophys. J. Lett.} \textbf{\bibinfo{volume}{528}}, \bibinfo{pages}{L13} (\bibinfo{year}{2000}), \eprint{astro-ph/9912263}.

\bibitem[{\citenamefont{de~Vries}(1999)}]{deVries:1999tiy}
\bibinfo{author}{\bibfnamefont{A.}~\bibnamefont{de~Vries}}, \bibinfo{journal}{Class. Quant. Grav.} \textbf{\bibinfo{volume}{17}}, \bibinfo{pages}{123} (\bibinfo{year}{1999}).

\bibitem[{\citenamefont{Akiyama et~al.}(2019{\natexlab{a}})}]{EventHorizonTelescope:2019dse}
\bibinfo{author}{\bibfnamefont{K.}~\bibnamefont{Akiyama}} \bibnamefont{et~al.} (\bibinfo{collaboration}{Event Horizon Telescope}), \bibinfo{journal}{Astrophys. J. Lett.} \textbf{\bibinfo{volume}{875}}, \bibinfo{pages}{L1} (\bibinfo{year}{2019}{\natexlab{a}}), \eprint{1906.11238}.

\bibitem[{\citenamefont{Akiyama et~al.}(2019{\natexlab{b}})}]{EventHorizonTelescope:2019ths}
\bibinfo{author}{\bibfnamefont{K.}~\bibnamefont{Akiyama}} \bibnamefont{et~al.} (\bibinfo{collaboration}{Event Horizon Telescope}), \bibinfo{journal}{Astrophys. J. Lett.} \textbf{\bibinfo{volume}{875}}, \bibinfo{pages}{L4} (\bibinfo{year}{2019}{\natexlab{b}}), \eprint{1906.11241}.

\bibitem[{\citenamefont{Akiyama et~al.}(2022{\natexlab{a}})}]{EventHorizonTelescope:2022xqj}
\bibinfo{author}{\bibfnamefont{K.}~\bibnamefont{Akiyama}} \bibnamefont{et~al.} (\bibinfo{collaboration}{Event Horizon Telescope}), \bibinfo{journal}{Astrophys. J. Lett.} \textbf{\bibinfo{volume}{930}}, \bibinfo{pages}{L17} (\bibinfo{year}{2022}{\natexlab{a}}), \eprint{2311.09484}.

\bibitem[{\citenamefont{Akiyama et~al.}(2022{\natexlab{b}})}]{EventHorizonTelescope:2022wkp}
\bibinfo{author}{\bibfnamefont{K.}~\bibnamefont{Akiyama}} \bibnamefont{et~al.} (\bibinfo{collaboration}{Event Horizon Telescope}), \bibinfo{journal}{Astrophys. J. Lett.} \textbf{\bibinfo{volume}{930}}, \bibinfo{pages}{L12} (\bibinfo{year}{2022}{\natexlab{b}}), \eprint{2311.08680}.

\bibitem[{\citenamefont{Akiyama et~al.}(2022{\natexlab{c}})}]{EventHorizonTelescope:2022wok}
\bibinfo{author}{\bibfnamefont{K.}~\bibnamefont{Akiyama}} \bibnamefont{et~al.} (\bibinfo{collaboration}{Event Horizon Telescope}), \bibinfo{journal}{Astrophys. J. Lett.} \textbf{\bibinfo{volume}{930}}, \bibinfo{pages}{L14} (\bibinfo{year}{2022}{\natexlab{c}}), \eprint{2311.09479}.

\bibitem[{\citenamefont{Virbhadra}(2022{\natexlab{a}})}]{Virbhadra:2022ybp}
\bibinfo{author}{\bibfnamefont{K.~S.} \bibnamefont{Virbhadra}}, \bibinfo{journal}{arXiv preprint}  (\bibinfo{year}{2022}{\natexlab{a}}), \eprint{2204.01792}.

\bibitem[{\citenamefont{Virbhadra}(2022{\natexlab{b}})}]{Virbhadra:2022iiy}
\bibinfo{author}{\bibfnamefont{K.~S.} \bibnamefont{Virbhadra}}, \bibinfo{journal}{Phys. Rev. D} \textbf{\bibinfo{volume}{106}}, \bibinfo{pages}{064038} (\bibinfo{year}{2022}{\natexlab{b}}), \eprint{2204.01879}.

\bibitem[{\citenamefont{Virbhadra and Ellis}(2000)}]{Virbhadra:1999nm}
\bibinfo{author}{\bibfnamefont{K.~S.} \bibnamefont{Virbhadra}} \bibnamefont{and} \bibinfo{author}{\bibfnamefont{G.~F.~R.} \bibnamefont{Ellis}}, \bibinfo{journal}{Phys. Rev. D} \textbf{\bibinfo{volume}{62}}, \bibinfo{pages}{084003} (\bibinfo{year}{2000}), \eprint{astro-ph/9904193}.

\bibitem[{\citenamefont{Virbhadra and Ellis}(2002)}]{Virbhadra:2002ju}
\bibinfo{author}{\bibfnamefont{K.~S.} \bibnamefont{Virbhadra}} \bibnamefont{and} \bibinfo{author}{\bibfnamefont{G.~F.~R.} \bibnamefont{Ellis}}, \bibinfo{journal}{Phys. Rev. D} \textbf{\bibinfo{volume}{65}}, \bibinfo{pages}{103004} (\bibinfo{year}{2002}).

\bibitem[{\citenamefont{Virbhadra}(2009)}]{Virbhadra:2008ws}
\bibinfo{author}{\bibfnamefont{K.~S.} \bibnamefont{Virbhadra}}, \bibinfo{journal}{Phys. Rev. D} \textbf{\bibinfo{volume}{79}}, \bibinfo{pages}{083004} (\bibinfo{year}{2009}), \eprint{0810.2109}.

\bibitem[{\citenamefont{Nascimento et~al.}(2020)\citenamefont{Nascimento, Petrov, Porfirio, and Soares}}]{Nascimento:2020ime}
\bibinfo{author}{\bibfnamefont{J.~R.} \bibnamefont{Nascimento}}, \bibinfo{author}{\bibfnamefont{A.~Y.} \bibnamefont{Petrov}}, \bibinfo{author}{\bibfnamefont{P.~J.} \bibnamefont{Porfirio}}, \bibnamefont{and} \bibinfo{author}{\bibfnamefont{A.~R.} \bibnamefont{Soares}}, \bibinfo{journal}{Phys. Rev. D} \textbf{\bibinfo{volume}{102}}, \bibinfo{pages}{044021} (\bibinfo{year}{2020}), \eprint{2005.13096}.

\bibitem[{\citenamefont{Furtado et~al.}(2021)\citenamefont{Furtado, Nascimento, Petrov, Porf\'\i{}rio, and Soares}}]{Furtado:2020puz}
\bibinfo{author}{\bibfnamefont{C.}~\bibnamefont{Furtado}}, \bibinfo{author}{\bibfnamefont{J.~R.} \bibnamefont{Nascimento}}, \bibinfo{author}{\bibfnamefont{A.~Y.} \bibnamefont{Petrov}}, \bibinfo{author}{\bibfnamefont{P.~J.} \bibnamefont{Porf\'\i{}rio}}, \bibnamefont{and} \bibinfo{author}{\bibfnamefont{A.~R.} \bibnamefont{Soares}}, \bibinfo{journal}{Phys. Rev. D} \textbf{\bibinfo{volume}{103}}, \bibinfo{pages}{044047} (\bibinfo{year}{2021}), \eprint{2010.11452}.

\bibitem[{\citenamefont{Kuang and \"Ovg\"un}(2022)}]{kuang2022strong}
\bibinfo{author}{\bibfnamefont{X.}~\bibnamefont{Kuang}} \bibnamefont{and} \bibinfo{author}{\bibfnamefont{A.}~\bibnamefont{\"Ovg\"un}}, \bibinfo{journal}{Annals Phys.} \textbf{\bibinfo{volume}{447}}, \bibinfo{pages}{169147} (\bibinfo{year}{2022}), \eprint{2205.11003}.

\bibitem[{\citenamefont{Gibbons and Werner}(2008)}]{Gibbons:2008rj}
\bibinfo{author}{\bibfnamefont{G.~W.} \bibnamefont{Gibbons}} \bibnamefont{and} \bibinfo{author}{\bibfnamefont{M.~C.} \bibnamefont{Werner}}, \bibinfo{journal}{Class. Quant. Grav.} \textbf{\bibinfo{volume}{25}}, \bibinfo{pages}{235009} (\bibinfo{year}{2008}), \eprint{0807.0854}.

\bibitem[{\citenamefont{Werner}(2012)}]{Werner_2012}
\bibinfo{author}{\bibfnamefont{M.~C.} \bibnamefont{Werner}}, \bibinfo{journal}{Gen. Rel. Grav.} \textbf{\bibinfo{volume}{44}}, \bibinfo{pages}{3047} (\bibinfo{year}{2012}), \eprint{1205.3876}.

\bibitem[{\citenamefont{Ishihara et~al.}(2016)\citenamefont{Ishihara, Suzuki, Ono, Kitamura, and Asada}}]{Ishihara:2016vdc}
\bibinfo{author}{\bibfnamefont{A.}~\bibnamefont{Ishihara}}, \bibinfo{author}{\bibfnamefont{Y.}~\bibnamefont{Suzuki}}, \bibinfo{author}{\bibfnamefont{T.}~\bibnamefont{Ono}}, \bibinfo{author}{\bibfnamefont{T.}~\bibnamefont{Kitamura}}, \bibnamefont{and} \bibinfo{author}{\bibfnamefont{H.}~\bibnamefont{Asada}}, \bibinfo{journal}{Phys. Rev. D} \textbf{\bibinfo{volume}{94}}, \bibinfo{pages}{084015} (\bibinfo{year}{2016}), \eprint{1604.08308}.

\bibitem[{\citenamefont{Ishihara et~al.}(2017)\citenamefont{Ishihara, Suzuki, Ono, and Asada}}]{Ishihara:2016sfv}
\bibinfo{author}{\bibfnamefont{A.}~\bibnamefont{Ishihara}}, \bibinfo{author}{\bibfnamefont{Y.}~\bibnamefont{Suzuki}}, \bibinfo{author}{\bibfnamefont{T.}~\bibnamefont{Ono}}, \bibnamefont{and} \bibinfo{author}{\bibfnamefont{H.}~\bibnamefont{Asada}}, \bibinfo{journal}{Phys. Rev. D} \textbf{\bibinfo{volume}{95}}, \bibinfo{pages}{044017} (\bibinfo{year}{2017}), \eprint{1612.04044}.

\bibitem[{\citenamefont{Ono et~al.}(2017)\citenamefont{Ono, Ishihara, and Asada}}]{Ono:2017pie}
\bibinfo{author}{\bibfnamefont{T.}~\bibnamefont{Ono}}, \bibinfo{author}{\bibfnamefont{A.}~\bibnamefont{Ishihara}}, \bibnamefont{and} \bibinfo{author}{\bibfnamefont{H.}~\bibnamefont{Asada}}, \bibinfo{journal}{Phys. Rev. D} \textbf{\bibinfo{volume}{96}}, \bibinfo{pages}{104037} (\bibinfo{year}{2017}), \eprint{1704.05615}.

\bibitem[{\citenamefont{Pantig and \"Ovg\"un}(2022{\natexlab{a}})}]{Pantig:2022toh}
\bibinfo{author}{\bibfnamefont{R.~C.} \bibnamefont{Pantig}} \bibnamefont{and} \bibinfo{author}{\bibfnamefont{A.}~\bibnamefont{\"Ovg\"un}}, \bibinfo{journal}{Eur. Phys. J. C} \textbf{\bibinfo{volume}{82}}, \bibinfo{pages}{391} (\bibinfo{year}{2022}{\natexlab{a}}), \eprint{2201.03365}.

\bibitem[{\citenamefont{Pantig and \"Ovg\"un}(2022{\natexlab{b}})}]{Pantig:2022whj}
\bibinfo{author}{\bibfnamefont{R.~C.} \bibnamefont{Pantig}} \bibnamefont{and} \bibinfo{author}{\bibfnamefont{A.}~\bibnamefont{\"Ovg\"un}}, \bibinfo{journal}{JCAP} \textbf{\bibinfo{volume}{08}}, \bibinfo{pages}{056} (\bibinfo{year}{2022}{\natexlab{b}}), \eprint{2202.07404}.

\bibitem[{\citenamefont{\"Ovg\"un et~al.}(2023{\natexlab{b}})\citenamefont{\"Ovg\"un, Pantig, and Rinc\'on}}]{Ovgun:2023ego}
\bibinfo{author}{\bibfnamefont{A.}~\bibnamefont{\"Ovg\"un}}, \bibinfo{author}{\bibfnamefont{R.~C.} \bibnamefont{Pantig}}, \bibnamefont{and} \bibinfo{author}{\bibfnamefont{A.}~\bibnamefont{Rinc\'on}}, \bibinfo{journal}{Eur. Phys. J. Plus} \textbf{\bibinfo{volume}{138}}, \bibinfo{pages}{192} (\bibinfo{year}{2023}{\natexlab{b}}), \eprint{2303.01696}.

\bibitem[{\citenamefont{Li and \"Ovg\"un}(2020)}]{Li:2020dln}
\bibinfo{author}{\bibfnamefont{Z.}~\bibnamefont{Li}} \bibnamefont{and} \bibinfo{author}{\bibfnamefont{A.}~\bibnamefont{\"Ovg\"un}}, \bibinfo{journal}{Phys. Rev. D} \textbf{\bibinfo{volume}{101}}, \bibinfo{pages}{024040} (\bibinfo{year}{2020}), \eprint{2001.02074}.

\bibitem[{\citenamefont{Li et~al.}(2020{\natexlab{a}})\citenamefont{Li, Zhang, and \"Ovg\"un}}]{Li:2020wvn}
\bibinfo{author}{\bibfnamefont{Z.}~\bibnamefont{Li}}, \bibinfo{author}{\bibfnamefont{G.}~\bibnamefont{Zhang}}, \bibnamefont{and} \bibinfo{author}{\bibfnamefont{A.}~\bibnamefont{\"Ovg\"un}}, \bibinfo{journal}{Phys. Rev. D} \textbf{\bibinfo{volume}{101}}, \bibinfo{pages}{124058} (\bibinfo{year}{2020}{\natexlab{a}}), \eprint{2006.13047}.

\bibitem[{\citenamefont{Wald}(1993)}]{Wald:1993nt}
\bibinfo{author}{\bibfnamefont{R.~M.} \bibnamefont{Wald}}, \bibinfo{journal}{Phys. Rev. D} \textbf{\bibinfo{volume}{48}}, \bibinfo{pages}{R3427} (\bibinfo{year}{1993}), \eprint{gr-qc/9307038}.

\bibitem[{\citenamefont{Donoghue}(1994)}]{Donoghue:1994dn}
\bibinfo{author}{\bibfnamefont{J.~F.} \bibnamefont{Donoghue}}, \bibinfo{journal}{Phys. Rev. D} \textbf{\bibinfo{volume}{50}}, \bibinfo{pages}{3874} (\bibinfo{year}{1994}), \eprint{gr-qc/9405057}.

\bibitem[{\citenamefont{Knorr et~al.}(2019)\citenamefont{Knorr, Ripken, and Saueressig}}]{Knorr:2019atm}
\bibinfo{author}{\bibfnamefont{B.}~\bibnamefont{Knorr}}, \bibinfo{author}{\bibfnamefont{C.}~\bibnamefont{Ripken}}, \bibnamefont{and} \bibinfo{author}{\bibfnamefont{F.}~\bibnamefont{Saueressig}}, \bibinfo{journal}{Class. Quant. Grav.} \textbf{\bibinfo{volume}{36}}, \bibinfo{pages}{234001} (\bibinfo{year}{2019}), \eprint{1907.02903}.

\bibitem[{\citenamefont{Donoghue and El-Menoufi}(2015)}]{Donoghue:2015nba}
\bibinfo{author}{\bibfnamefont{J.~F.} \bibnamefont{Donoghue}} \bibnamefont{and} \bibinfo{author}{\bibfnamefont{B.~K.} \bibnamefont{El-Menoufi}}, \bibinfo{journal}{JHEP} \textbf{\bibinfo{volume}{10}}, \bibinfo{pages}{044} (\bibinfo{year}{2015}), \eprint{1507.06321}.

\bibitem[{\citenamefont{Barvinsky and Vilkovisky}(1990)}]{Barvinsky:1990up}
\bibinfo{author}{\bibfnamefont{A.~O.} \bibnamefont{Barvinsky}} \bibnamefont{and} \bibinfo{author}{\bibfnamefont{G.~A.} \bibnamefont{Vilkovisky}}, \bibinfo{journal}{Nucl. Phys. B} \textbf{\bibinfo{volume}{333}}, \bibinfo{pages}{471} (\bibinfo{year}{1990}).

\bibitem[{\citenamefont{Deser and Redlich}(1986)}]{Deser:1986xr}
\bibinfo{author}{\bibfnamefont{S.}~\bibnamefont{Deser}} \bibnamefont{and} \bibinfo{author}{\bibfnamefont{A.~N.} \bibnamefont{Redlich}}, \bibinfo{journal}{Phys. Lett. B} \textbf{\bibinfo{volume}{176}}, \bibinfo{pages}{350} (\bibinfo{year}{1986}), \bibinfo{note}{[Erratum: Phys.Lett.B 186, 461 (1987)]}.

\bibitem[{\citenamefont{Asorey et~al.}(1997)\citenamefont{Asorey, Lopez, and Shapiro}}]{Asorey:1996hz}
\bibinfo{author}{\bibfnamefont{M.}~\bibnamefont{Asorey}}, \bibinfo{author}{\bibfnamefont{J.~L.} \bibnamefont{Lopez}}, \bibnamefont{and} \bibinfo{author}{\bibfnamefont{I.~L.} \bibnamefont{Shapiro}}, \bibinfo{journal}{Int. J. Mod. Phys. A} \textbf{\bibinfo{volume}{12}}, \bibinfo{pages}{5711} (\bibinfo{year}{1997}), \eprint{hep-th/9610006}.

\bibitem[{\citenamefont{Teixeira et~al.}(2020)\citenamefont{Teixeira, Shapiro, and Ribeiro}}]{Teixeira:2020kew}
\bibinfo{author}{\bibfnamefont{P.~d.~M.} \bibnamefont{Teixeira}}, \bibinfo{author}{\bibfnamefont{I.~L.} \bibnamefont{Shapiro}}, \bibnamefont{and} \bibinfo{author}{\bibfnamefont{T.~G.} \bibnamefont{Ribeiro}}, \bibinfo{journal}{Grav. Cosmol.} \textbf{\bibinfo{volume}{26}}, \bibinfo{pages}{185} (\bibinfo{year}{2020}), \eprint{2003.04503}.

\bibitem[{\citenamefont{Deser and van Nieuwenhuizen}(1974)}]{Deser:1974cz}
\bibinfo{author}{\bibfnamefont{S.}~\bibnamefont{Deser}} \bibnamefont{and} \bibinfo{author}{\bibfnamefont{P.}~\bibnamefont{van Nieuwenhuizen}}, \bibinfo{journal}{Phys. Rev. D} \textbf{\bibinfo{volume}{10}}, \bibinfo{pages}{401} (\bibinfo{year}{1974}).

\bibitem[{\citenamefont{Barvinsky and Vilkovisky}(1983)}]{Barvinsky:1984jd}
\bibinfo{author}{\bibfnamefont{A.~O.} \bibnamefont{Barvinsky}} \bibnamefont{and} \bibinfo{author}{\bibfnamefont{G.~A.} \bibnamefont{Vilkovisky}}, \bibinfo{journal}{Phys. Lett. B} \textbf{\bibinfo{volume}{131}}, \bibinfo{pages}{313} (\bibinfo{year}{1983}).

\bibitem[{\citenamefont{Fursaev}(1995)}]{Fursaev:1994te}
\bibinfo{author}{\bibfnamefont{D.~V.} \bibnamefont{Fursaev}}, \bibinfo{journal}{Phys. Rev. D} \textbf{\bibinfo{volume}{51}}, \bibinfo{pages}{5352} (\bibinfo{year}{1995}), \eprint{hep-th/9412161}.

\bibitem[{\citenamefont{El-Menoufi}(2016)}]{El-Menoufi:2015cqw}
\bibinfo{author}{\bibfnamefont{B.~K.} \bibnamefont{El-Menoufi}}, \bibinfo{journal}{JHEP} \textbf{\bibinfo{volume}{05}}, \bibinfo{pages}{035} (\bibinfo{year}{2016}), \eprint{1511.08816}.

\bibitem[{\citenamefont{El-Menoufi}(2017)}]{El-Menoufi:2017kew}
\bibinfo{author}{\bibfnamefont{B.~K.} \bibnamefont{El-Menoufi}}, \bibinfo{journal}{JHEP} \textbf{\bibinfo{volume}{08}}, \bibinfo{pages}{068} (\bibinfo{year}{2017}), \eprint{1703.10178}.

\bibitem[{\citenamefont{Goroff and Sagnotti}(1986)}]{Goroff:1985th}
\bibinfo{author}{\bibfnamefont{M.~H.} \bibnamefont{Goroff}} \bibnamefont{and} \bibinfo{author}{\bibfnamefont{A.}~\bibnamefont{Sagnotti}}, \bibinfo{journal}{Nucl. Phys. B} \textbf{\bibinfo{volume}{266}}, \bibinfo{pages}{709} (\bibinfo{year}{1986}).

\bibitem[{\citenamefont{Perlick and Tsupko}(2022)}]{Perlick:2021aok}
\bibinfo{author}{\bibfnamefont{V.}~\bibnamefont{Perlick}} \bibnamefont{and} \bibinfo{author}{\bibfnamefont{O.~Y.} \bibnamefont{Tsupko}}, \bibinfo{journal}{Phys. Rept.} \textbf{\bibinfo{volume}{947}}, \bibinfo{pages}{1} (\bibinfo{year}{2022}), \eprint{2105.07101}.

\bibitem[{\citenamefont{Ruffini}(1813)}]{ruffini1813riflessioni}
\bibinfo{author}{\bibfnamefont{P.}~\bibnamefont{Ruffini}}, \emph{\bibinfo{title}{Riflessioni intorno alla soluzione delle equazioni algebraiche generali opuscolo del cav. dott. Paolo Ruffini...}} (\bibinfo{publisher}{presso la Societa Tipografica}, \bibinfo{year}{1813}).

\bibitem[{\citenamefont{King}(2015)}]{King_2015}
\bibinfo{author}{\bibfnamefont{A.}~\bibnamefont{King}}, \bibinfo{journal}{Monthly Notices of the Royal Astronomical Society: Letters} \textbf{\bibinfo{volume}{456}}, \bibinfo{pages}{L109–L112} (\bibinfo{year}{2015}), ISSN \bibinfo{issn}{1745-3933}, \urlprefix\url{http://dx.doi.org/10.1093/mnrasl/slv186}.

\bibitem[{\citenamefont{Li et~al.}(2020{\natexlab{b}})\citenamefont{Li, He, and Zhou}}]{Li:2019vhp}
\bibinfo{author}{\bibfnamefont{Z.}~\bibnamefont{Li}}, \bibinfo{author}{\bibfnamefont{G.}~\bibnamefont{He}}, \bibnamefont{and} \bibinfo{author}{\bibfnamefont{T.}~\bibnamefont{Zhou}}, \bibinfo{journal}{Phys. Rev. D} \textbf{\bibinfo{volume}{101}}, \bibinfo{pages}{044001} (\bibinfo{year}{2020}{\natexlab{b}}), \eprint{1908.01647}.

\bibitem[{\citenamefont{Gibbons}(2016)}]{Gibbons:2015qja}
\bibinfo{author}{\bibfnamefont{G.~W.} \bibnamefont{Gibbons}}, \bibinfo{journal}{Class. Quant. Grav.} \textbf{\bibinfo{volume}{33}}, \bibinfo{pages}{025004} (\bibinfo{year}{2016}), \eprint{1508.06755}.

\end{thebibliography}
\bibliographystyle{apsrev}

\end{document}